\documentclass[a4paper]{article}
\usepackage{lipsum}
\usepackage[noadjust, nobreak]{cite}
\usepackage{amsmath}
\usepackage{floatrow}
\usepackage{layout}
\usepackage{amssymb} 
\usepackage{multirow}
\usepackage{authblk}
\usepackage{indentfirst}
\usepackage{bm}
\usepackage[sort]{natbib}
\usepackage{algorithm} 
\usepackage[noend]{algpseudocode}
\usepackage{amsmath}
\algnewcommand\algorithmicforeach{\textbf{for each}}
\algdef{S}[FOR]{ForEach}[1]{\algorithmicforeach\ #1\ \algorithmicdo}
\newcommand{\comb}[2]{{C}_{#2}^{#1}}
\usepackage{caption}
\usepackage{graphicx}
\usepackage{tikz}
\usetikzlibrary{shapes.geometric, arrows}
\usepackage{mathtools}
\floatstyle{plaintop}
\restylefloat{table}
\usepackage{graphicx}
\usepackage[flushleft]{threeparttable}
\usepackage{subcaption}

\begin{document}
\title{\bf
	A Two-Stage Variable Selection Approach for Correlated High Dimensional Predictors
}
\author{Zhiyuan Li* \\ Email: li3z3@mail.uc.edu}
\affil{Department of EECS, University of Cincinnati}
\date{}
\maketitle
\begin{abstract}
		When fitting statistical models, some predictors are often found to be correlated with each other, and functioning together. Many group variable selection methods are developed to select the groups of predictors that are closely related with the continuous or categorical response. These existing methods usually assume the group structures are well known. For example, variables with similar practical meaning, or dummy variables created by a categorical data. However, in practice, it is impractical to know the exact group structure, especially when the variable dimensional is large. As a result, the group variable selection results may be affected. To solve the challenge, we propose a two-stage approach which combines a variable clustering stage and a group variable stage for the group variable selection problem. The variable clustering stage uses information from the data to find a group structure, which improves the performance of the existing group variable selection methods. For ultrahigh dimensional data, where the predictors are much larger than observations, we incorporated a variable screening method in the first stage and shows the advantages of such approach. In this article, we compared and discussed the performance of four existing group variable selection methods under different simulation models, with and without the variable clustering stage. The two-stage method shows a better performance, in terms of the prediction accuracy, as well as in the accuracy to select active predictors. An athlete data is also used to show the advantages of the proposed method
\end{abstract}

\section{Introduction}
	    Regression and classification, two highly utilized methods in supervise learning, are conducted to perform a functional relationship between response and predictors (or variables). They are widely applied  predicting the new observation outcomes and to select the important variables. When building a regression or classification model, a simpler model, in other words, a model with less variables, is easier to be interpreted. In addition, removing unimportant variables can also reduce the model loss, risk of over fitting and further improve the prediction performance. Variables selection is a method to explore the important variables that related the response, it has been applying in both industry and research institute. In usual regression set up, we assume each variables are independent and the general linear regression model is defined by 
	    \numberwithin{equation}{section}
    	\begin{align} \label{eq:1.1}
    	\bm{Y}&=\bm{X}\bm{\beta}+\bm{\epsilon}
    	\end{align}
    	where $\bm{Y} \in R^{n\times 1}$ is the response vector, $\bm{X}$ is the predictor matrix corresponding to the p predictors. Without loss of generality, we assume $\bm{Y}$ and $\bm{X}$ are centered. $\bm{\beta} \in R^{p\times 1}$ is the coefficient vector, $\beta_i\in \bm{\beta}$ and $\bm{\epsilon} \in R^{n\times 1}$ is a random error with mean $\bm{0}$ and variance $\sigma^2\bm{I}$. The ordinary least squares (OLS) method estimates $\bm{\beta}$ by minimizing the residual sum square ($RSS$) as follows
    	\begin{align}
    	\label{eq:1.2}
    	\hat{\bm{\beta}}=\underset{\beta}{\operatorname{argmin}}\{||\bm{Y}-\bm{X}\bm{\beta}||^2_{2}\} 
    	\end{align}
        where $L(\bm{\beta})=||\bm{Y}-\textbf{X}\bm{\beta}||^2_{2}$ is the loss function, and $||\cdot||_2$ stands for the $L_2$ norm. Based on function \eqref{eq:1.2}, an unbiased estimator $\hat{\bm{\beta}}$ can be solved for $\bm{\beta}$, such that, $\hat{\bm{\beta}}=(\bm{X}^T \bm{X})^{-1} \bm{X}^T \bm{Y}$. For binary classification problem, let $\bm{X_i}=(X_{11},X_{12},\cdots,X_{1p})^T\subseteq \bm{X}$, where $\bm{X}=(\bm{X_1},\bm{X_2},\cdots,\bm{X_n})^T$ for $i=1,2,\cdots n$, and $Y_i \in \{0,1\}\sim$ Binomial$(n,p)$ and $Y_i \in Y$, where $p[Y_i=1|\bm{X_i}]$ is the probability that the response belongs to the category 1 for $i$th observation. Hence, the logistic model is defined as
        \begin{align} \label{eq:1.3}
            log(\frac{p[Y_i=1|\bm{X}_i]}{1-p[Y_i=1|\bm{X}_i]})=\bm{X}_i^T\bm{\beta}\enspace \text{for}\enspace i=1,2,\cdots,n
        \end{align}
        where the coefficient estimator $\hat{\bm{\beta}}$ is estimated by minimizing the loss function, which is defined as 
        \begin{align} \label{eq:1.4}
            L(\bm{\beta})&=-\frac{1}{n}log\{\prod_{i=1}^n [p(Y_i=1|\bm{X}_i)]^{Y_i}[1-p(Y_i=1|\bm{X}_i)]^{1-Y_i}\}
        \end{align}
        and the Gradient Descent or Newton’s Method can be applied to solve the optimal solution for $\hat{\bm{\beta}}$ in function \eqref{eq:1.4}.
        
        In practice, $\hat{\bm{\beta}}$ trends to be inaccurate when the variables are correlated to each other. Therefore, selecting the important variables in equations \eqref{eq:1.1} and \eqref{eq:1.3} becomes a majority problem. It is the same as deciding which coefficients should be set to zero in $\bm{\beta}$. A simple approach is to make a hypothesis test for each variable by producing an ANOVA table and eliminate the insignificant variables. However, ANOVA is valid only when the variables are orthogonal, and it does not essentially solve the multicollinearity issue. Several traditional approaches, such as the best subset regression, which was proposed by \citet{hocking1967selection}, can be applied. In the best subset regression, some popular model selection criterion, such as Akaike Information Criterion (AIC), Bayesian Information Criterion (BIC), Coefficient of Determination ($R^2$), have been frequently applied to evaluate each candidate model and choose the best model associated with the optimal score. The best subset regression performs very accurate but it only works well when the number of predictors is relatively small. The reason is that, the time complexity for the best subset regression is $O(2^n)$, and it increasingly cost a big amount of computational work with the number of predictors increases. Another popular variable selection method is stepwise methods, for example, the forward selection or backward elimination, which was proposed by \citet{efroymson1966stepwise}. It has been applied widely to select the individual variables due to its great performance criterion in terms of accuracy and fast computation. However, the stepwise approach was classified to greedy algorithms, and it trends to attaches local optimizer instead of the global optimizer.
        
        In modern statistics studies, with the fast growth of computing power on storage of database and data warehouse, massive and ultrahigh dimensional datasets normally exist in various fields, such as social media, economics, and medicine. Several recent studies (\citet{hoerl1970ridge}, \citet{foster1994risk}, \citet{tibshirani1996regression}, \citet{efron2004least}, \citet{zou2005regularization}) have been conducted and showed that the traditional best subset regression methods are lack of stability, especially when the number of predictors are greater than the number of observations. \citet{hoerl1970ridge} proposed the Ridge and introduced the shrinkage idea. \citet{tibshirani1996regression} proposed the famous Lasso to perform variable selection and it displayed great performance on both estimation accuracy and computation efficiency. After Lasso had been introduced, many methods follow similar shrinkage idea were also developed. For example, \citet{fan2001variable} proposed the Smoothly Clipped Absolute Deviation (SCAD) and \citet{zhang2010nearly} proposed the Minimax Concave Penalty (MCP) to solve the Oracle properties issue in Lasso. Meanwhile, \citet{zou2005regularization} proposed the Elastic Net approach and overcame the disadvantage that Lasso, SCAD and MCP all ignore the correlation between the predictors.
        
        In high dimensional data, the predictors usually are correlated with each other and functioning together in groups. The groups are usually called factors or components. Although individual variable selection methods like Lasso, SCAD and MCP achieve excellent variable selection performance, for variables with a group structure, they will usually randomly choose one of them and penalize all the other coefficients to zero. This characteristic can provide simple and sparse solution, however, it is unfavorable,  since intuitively one want to keep all the predictors inside a group active if they functioning together. To solve this problem, \citet{yuan2006model} proposed a group variable selection method that is called group Lasso. It penalizes the group coefficients based on the Lasso penalty. This method assumes the group structure of the predictors are known in advance. Follow similar idea, extended SCAD and MCP penalties and proposed group SCAD and group MCP methods. In addition, considering the possibility that only few predictors in a group are truly active, \citet{simon2013sparse} proposed the sparse group Lasso method, which adds an additional penalty term based on group Lasso, and this allow sparsity on both groups and individual variables with a group.
        
        A regular challenge for the current existing group variable selection models is that the researchers need to have a prior knowledge of group structure among variables. Such information is usually obtained based on prior research in the certain field, data are collected in groups, or variables are dummy variables of a categorical predictor. Therefore, researchers usually need to pre-define which variables should be defined as a group. However, it is too subjective sometimes to determine the group structure based on how data is collected, since the variables may have other unknown relationship. It is also impractical to know the grouping information if the predictor dimensional is quite large. In addition, in many research projects, one of the goals of the analysis is to discover the relationship and group structures among the predictors and see how they are related with the response. Therefore, it is not possible to know a valid group information in advance. Due to above mentioned issues, letting the data itself show the group structure seems make sense. In other words, we can find the group structure before fitting a group variable selection model. This idea was introduced by \citet{park2007averaged}, who proposed a method that combines hierarchical clustering and Lasso to solve the challenge of having more predictors than observations, \citet{tolocsi2011classification} also applied the clustering method, which refer to \citet{park2007averaged}, for classification problem in high correlation of genomics dataset. In group regularization model, \citet{buhlmann2013correlated} proposed the cluster group Lasso (CGL) to identify the group structure for group Lasso by utilizing the noveland bottom-up agglomerative clustering algorithm based on canonical correlations. \citet{gauraha2017pre} introduced a pre-selection method using elastic net + CGL for high dimension variable space. A further challenge with the group clustering is that both continuous and categorical predictors exist, a method that can simultaneously deal with these two types of variables is preferred.  
        
        In this paper, we propose a two-stage group variable selection approach based on variable clustering to select the optimal group structure in the first stage and group variable selection as the second stage. For dataset with much higher number of predictors than the observations, we edit the first stage by applying a variable screening method first to quickly get a subset containing active predictors, then apply the regular grouping algorithm. Another goal of this paper is to compare the performance of various group variable selection methods for both continuous and categorical responses, under various model set up. The paper is organized as follows: Section 2 introduces variable selection, group/sparse group variable selection methods, variable clustering and variable screening algorithms. The proposed two-stage group variable selection method is introduced and discussed in Section 3. Simulation examples to compare different group variable selection methods as well as the proposed approach are included in Section 4. Section 5 contains an example of real data analysis. Discussion and conclusions are in Section 6.

        \section{Method}
            \subsection{Individual Variable Selection for Regression}
            We first overview some existing variable selection methods for regression and classification. Consider a regression model as in equation \eqref{eq:1.1}, the following methods follow the idea to penalize the regression coefficients based on different penalty functions. 
            
            \subsubsection{Lasso}
            The Lasso is proposed by \citet{tibshirani1996regression} to perform variable selection and parameter estimation, and the estimator is defined as follows:
            \begin{align}\label{eq:2.1}
                \hat{\bm{\beta}}^{\text{Lasso}}(\lambda)=\underset{\beta}{\operatorname{argmin}}\{||\bm{Y}-\bm{X}\bm{\beta}||^2_{2}+\sum_{i=1}^p\lambda ||\beta_i||_1\}
            \end{align}
             where $\lambda\ge 0$ is a tuning parameter, $||\cdot||_1$ stands for the $L_1$ norm and $\textbf{X}$ is a centered and scaled matrix. When variables are strongly correlated, Lasso regression likely selects one variable from the correlated variables, and coefficients for the remaining variables are decayed to 0. The sparsity in the solution is introduced by the $L_1$ norm. If the $\lambda$ in function \eqref{eq:2.1} is larger, the more coefficients in the Lasso estimator $\hat{\bm{\beta}}^{\text{Lasso}}(\lambda)$ will be shrunk to 0.  What is more, if $\lambda \to \infty$, then $\hat{\bm{\beta}}^{\text{Lasso}} \to 0$, and it leads to an empty or null model. The most popular solutions for estimating parameters in Lasso is the Least Angle Regression (LARS; \cite{efron2004least}) and Coordinate Descent (\citet{friedman2010regularization})
             
            \subsubsection{SCAD and MCP}
            To adjust the excessive shrinkage power based on $L_1$ norm for Lasso, \citet{fan2001variable} proposed a non-convex function, Smoothly Clipped Absolute Deviation (SCAD), with the estimator defined as 
            \begin{align} \label{eq:2.2}
                 \hat{\bm{\beta}}^{\text{SCAD}}(\lambda,\gamma)=\underset{\beta}{\operatorname{argmin}}\{||\bm{Y}-\bm{X}\bm{\beta}||^2_{2}+\sum_{i=1}^{p}P_{\lambda,\gamma}^{\text{SCAD}}(||\beta_i||_1)\}
            \end{align}
            with the penalty function $P_{\lambda,\gamma}^{\text{SCAD}}(\cdot)$, such that
            \begin{equation*}
            P_{\lambda,\gamma}^{\text{SCAD}}(w) = \begin{cases}
            \lambda |w|, & |w|\le \lambda;  \\
            \frac{2\gamma|w|-w^2-\lambda^2}{2(\gamma-1)}, & \lambda<|w|< \gamma\lambda;  \\
             \frac{(\gamma+1)\lambda^2}{2}, & |w|\ge\gamma\lambda;
            \end{cases}
            \end{equation*}
            where $\gamma>2$ and $\lambda>0$ are the tuning parameters, and Fan suggested $\gamma=3.7$. The penalty function from SCAD is a non-convex function and can strongly shrink the small regression coefficients but shrink the large coefficients weakly.
        
            The Minimax Concave Penalty (MCP) is another non-convex function, which was proposed by \citet{zhang2010nearly}, and it solved the estimation under the approximate unbiasedness and find a solution for concavity computational challenge with minimum penalty. The coefficient estimator for MCP penalty is defined as
            \begin{align} \label{eq:2.3}
            \hat{\bm{\beta}}^{\text{MCP}}(\lambda,\gamma)=\underset{\beta}{\operatorname{argmin}}\{||\bm{Y}-\bm{X}\bm{\beta}||^2_{2}+\sum_{i=1}^{p}P_{\lambda,\gamma}^{\text{MCP}}(||\beta_i||_1)\}
             \end{align}
            with the penalty function $P_{\lambda,\gamma}^{\text{MCP}}(\cdot)$, such that
            \begin{equation*}
            P_{\lambda,\gamma}^{\text{MCP}}(w) = \begin{cases}
            \lambda |w|-\frac{w^2}{2\gamma}, & |w|\le \gamma\lambda; \\
            \frac{\gamma\lambda^2}{2}, & |w|>\gamma\lambda;
            \end{cases}
            \end{equation*}
            where $\gamma>1$ and $\lambda>0$ are the tuning parameters. The idea of SCAD and MCP is very similar. Both SCAD and MCP enjoy a strong performance on Oracle properties in terms of on unbiasedness, sparsity and continuity, which the Lasso method does not achieve.
            
            Comparing between SCAD and MCP, we typically compare their derivative of the penalty, which are defined as follows 
            \begin{equation*}
            P_{\lambda,\gamma}^{'\text{SCAD}}(w)=
            \begin{cases}
            \lambda, & |w|\le \lambda;  \\
            \frac{\gamma \lambda-|w|}{\gamma-1}, & \lambda<|w|< \gamma\lambda;  \\
             0, & |w|\ge\gamma\lambda
            \end{cases} 
            \end{equation*}
            and 
            \begin{equation*}
            P_{\lambda,\gamma}^{'\text{MCP}}(w)=
            \begin{cases}
            (\lambda-\frac{|w|}{\gamma})\text{sign}(x), & |w|\le \gamma \lambda;  \\
             0, & |w|>\gamma\lambda
            \end{cases}
            \end{equation*}
            
            Figure 1 displays that SCAD and MCP starts using same penalization rate as Lasso's, then reduce the rate down to 0 as the size of coefficient increases. Besides, MCP starts to decrease the penalization rate immediately to 0 while SCAD keeps the same rate with the Lasso's before decreasing.
            \begin{figure}[H]
             \caption{Penalization Rate for Lasso, MCP, SCAD}
                \centering
                \includegraphics[width=12cm, height=8cm]{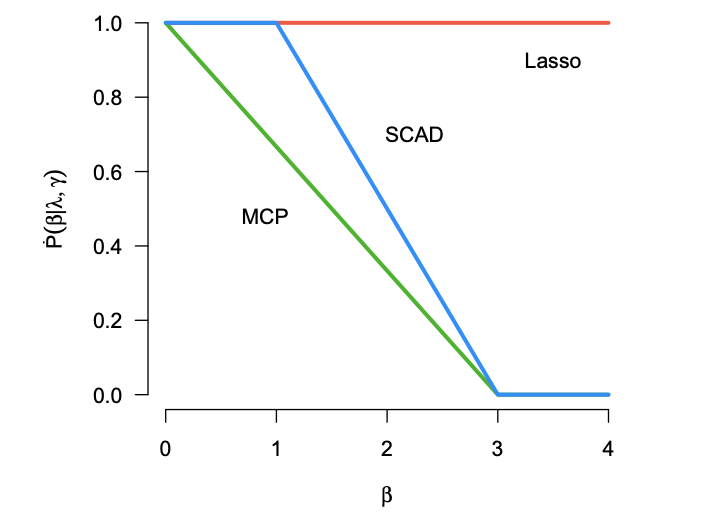}
            \end{figure}

            \subsection{Individual Variable Selection in Classification}
            For classification problem, we can apply the same penalty from Lasso, SCAD and MCP. The coefficient estimator $\hat{\beta}$ can be defined as 
            \begin{align*} \
                \hat{\bm{\beta}}^{\text{Lasso}}(\lambda)&=\underset{\beta}{\operatorname{argmin}}\{L(\bm{\beta})+\sum_{i=1}^p\lambda||\beta_i||_1\} \\
                \hat{\bm{\beta}}^{\text{SCAD}}(\lambda,\gamma)&=\underset{\beta}{\operatorname{argmin}}\{L(\bm{\beta})+ \sum_{i=1}^pP_{\lambda,\gamma}^{\text{SCAD}}(||\beta_i||_1)\} \\
                \hat{\bm{\beta}}^{\text{MCP}}(\lambda,\gamma)&=\underset{\beta}{\operatorname{argmin}}\{L(\bm{\beta})+ \sum_{i=1}^pP_{\lambda,\gamma}^{\text{MCP}}(||\beta_i||_1)\} \\
            \end{align*}
            where $L(\bm{\beta})$ is the loss function follows \eqref{eq:1.4}, parameters $\lambda$,$\gamma$ and penalty functions $P_{\lambda,\gamma}^{\text{SCAD}}(\cdot)$, $P_{\lambda,\gamma}^{\text{MCP}}(\cdot)$ enjoy the same properties in regression cases from section 2.1.

            \subsection{Group Variable Selection in Regression}
            For the predictors and response $\bm{X}$ and $\bm{Y}$, suppose the predictor variables are separated into $K$ groups. Thus, matrix $\bm{X}$ can be written as $(\bm{X}_1,\bm{X}_2, \cdots,\bm{X}_K)$, where  $\bm{X}_k=(X_{k1}, X_{k2}, \cdots,X_{kp_k})$, for $k=1,2,3...,K$, and $p_k$ be the group size for the $k$-th group. The corresponding ${p\times 1}$ coefficient vector $\bm{\beta}$ can thus be represented as $(\bm{\beta}_1,\bm{\beta}_2, \cdots,\bm{\beta}_K)$, with $\bm{\beta}_k^T=(\beta_{k1},\beta_{k2},\cdots,\beta_{kp_k})$. Suppose the ${n\times 1}$ error vector $\bm{\epsilon}\sim N(\bm{0},\sigma^2\bm{I})$, regression model with group structure can be rewritten as
            \begin{align}\label{eq:2.4}
            \bm{Y}=\sum_{k=1}^{K}\bm{X}_k\bm{\beta}_k+\bm{\epsilon}
            \end{align}
            
            \subsubsection{Group Lasso}
            The group lasso was proposed by \citet{yuan2006model}, it assumes that each group are orthonormal of the model matrices. The coefficient estimator of group Lasso is defined as
            \begin{align}\label{eq:2.5}
             \hat{\bm{\beta}}^{\text{grLasso}}(\lambda)= \underset{\beta}{\operatorname{argmin}}\{||\bm{Y}-\sum_{k=1}^K\bm{X}_k\bm{\beta}_k||_2^2+\lambda\sum_{k=1}^K\sqrt{p_k}||\bm{\beta}_k||_2\} 
            \end{align}
            where $\lambda\ge 0$ is a tuning parameter to control the number of groups. The function \eqref{eq:2.5} is a convex function so that the global optimizer solution exist. To find the solution for group Lasso, \citet{yuan2006model} proposed the group LARS algorithm based on LARS. Futhermore, \citet{breheny2009penalized} proposed the Locally Coordinate Decent (LCD) based on Coordinate Decent. Choosing a large value of $\lambda$ can screen out entire group variables while a low value of $\lambda$ can include all group variables into the model. In other words, if $\lambda \to \infty$, then $\hat{\bm{\beta}}^{\text{grLasso}} \to 0$, and leads to a null model; if $\lambda=0$, then $\hat{\bm{\beta}}^{\text{grLasso}} = \hat{\bm{\beta}}^{\text{OLS}}$.
            
            The group Lasso requires prior knowledge about the group structure, if each group only contains one variable, i.e.,$K=p$, then group Lasso will be reduced to Lasso; if there only exist one group, i.e, $K=1$, then group Lasso will be reduced to Ridge regression. If it reduces to ridge regression, group Lasso cannot screen out any variables, since the penalty function used $L_2$ norm on the subspace for each groups. Even though the group Lasso enjoys very good variable selection performance, as an extension of Lasso, group Lasso does not have Oracle properties. 
            
            \subsubsection{Group SCAD and Group MCP}
             Follow similar group variable selection idea as above , \citet{wang2007group} proposed the group SCAD, and \citet{huang2012selective} proposed the group MCP based on the SCAD and MCP penalty respectively, and the estimator is defined as follows 
             \begin{align}\label{eq:2.6}
                  \hat{\bm{\beta}}^{\text{grSCAD}}(\lambda,\gamma)&=\underset{\beta}{\operatorname{argmin}}\{||\bm{Y}-\sum_{k=1}^K\bm{X}_k\bm{\beta}_k||_2^2+\lambda \sum_{k=1}^KP_{\lambda,\gamma}^{\text{SCAD}}(||\bm{\beta}_k||_2)\} \\
                  \hat{\bm{\beta}}^{\text{grMCP}}(\lambda,\gamma)&=\underset{\beta}{\operatorname{argmin}}\{||\bm{Y}-\sum_{k=1}^K\bm{X}_k\bm{\beta}_k||_2^2+\lambda \sum_{k=1}^KP_{\lambda,\gamma}^{\text{MCP}}(||\bm{\beta}_k||_2)\} 
             \end{align}
              where both tuning parameters $\lambda, \gamma$ and penalty functions $P_{\lambda,\alpha}^{\text{SCAD}}(\cdot)$, $P_{\lambda,\alpha}^{\text{MCP}}(\cdot)$ are corresponding to the properties of SCAD and MCP, respectively. Both group SCAD and group MCP solved the weakness of biased estimation from group Lasso, and achieves the oracle properties.  
             
             \subsubsection{Sparse Group Variable Selection}
             In the above mentioned group variable selection models, the penalty function is computed by the sum of $L_2$ norm of group variables. Similar to Lasso, it can set the coefficient of some groups to 0, and similar to Ridge, it can only shrink the coefficients within a group but never set any one of them to 0. This indicates that individual variables within a group are not able to be selected for group variable selection methods. However, in practice, although there maybe high correlation among the variables within a group, only a few variables are truly related with the response. To allow group variable selection methods to select individual variables within some groups, the sparse group penalty has been introduced. This method combines two penalty function to achieve a double-layered variable selection, where one of the layer is to penalize the group variables and the other is to penalize individual variables within some groups. The general form of double-layered penalty function can be defined as 
             \begin{align}\label{eq:2.8}
             F_{\lambda}(\bm{\beta}) = \lambda_1\sum_{k=1}^{K}P_1(||\bm{\beta}_k||_2)
             +\lambda_2\sum_{k=1}^{K}\sum_{p_k=1}^{p_K}P_2(||\beta_{kp_k}||_1)
             \end{align}
             where $P_1(\cdot)$ is to control the group coefficient, and $P_2(\cdot)$ is to control the individual coefficient within some groups.
             
             \citet{simon2013sparse} proposed an extended version of group Lasso method, so-called sparse group Lasso. This method combined Lasso and group Lasso through adding an additional $L_1$ norm penalty from Lasso to each group. The coefficient estimator of the sparse group Lasso is defined as 
             \begin{align} \label{eq:2.9}
              \hat{\bm{\beta}}^{\text{SGL}}(\lambda)=\underset{\beta}{\operatorname{argmin}}\{||\bm{Y}-\sum_{k=1}^K\bm{X}_k||_2^2+\lambda_1\sum_{k=1}^K||\bm{\beta}_k||_2+\lambda_2||\bm{\beta}||_1\}
             \end{align}
              where $\lambda_1, \lambda_2 \ge 0$ are the tuning parameters. In equation \eqref{eq:2.9}, the first penalty term $\lambda_1\sum_{k=1}^K||\bm{\beta_k}||_2$ controls the group sparsity and the second penalty term $\lambda_2||\bm{\beta}||_1$ controls the variable sparsity within some certain groups. The penalty of sparse group Lasso can be considered as involving both penalty terms from group Lasso and Lasso, such that, if $\lambda_1 =0,\lambda_2>0$, then $\hat{\bm{\beta}}^{\text{SGL}}(\lambda)= \hat{\bm{\beta}}^{\text{Lasso}}(\lambda)$, and if $\lambda_1>0,\lambda_2=0$, then $\hat{\bm{\beta}}^{\text{SGL}}(\lambda)= \hat{\bm{\beta}}^{\text{grLasso}}(\lambda)$. Besides, equation \eqref{eq:2.9} is a convex function since it is sum of convex function with both Lasso and group Lasso penalties. Therefore, this method obtains a global optimizer and achieves efficient computation for its estimated coefficients. Because sparse group Lasso is combined by group Lasso and Lasso, where none of them has Oracle properties, thus, it also does not have Oracle properties. 
             
             \subsection{Group Variable Selection in Classification}
             For classification problems, for example, when the response vector $\bm{Y}$ is binary, the classification model with group structures can be rewritten as 
             \begin{align}\label{eq:2.10}
               log(\frac{p[Y_i=1|\bm{X}_i]}{1-p[Y_i=1|\bm{X}_i]})=\sum_{k=1}^{K}\bm{X}_{ik}^T\bm{\beta}_k
             \end{align}
             Similar to section 2.2, the coefficient estimators for group Lasso, group SCAD, group MCP and sparse group Lasso are defined as follows
             \begin{align*}
             \hat{\bm{\beta}}^{\text{grLasso}}(\lambda)&= \underset{\beta}{\operatorname{argmin}}\{L(\bm{\beta})+\lambda\sum_{k=1}^K\sqrt{p_k}||\bm{\beta}_k||_2\} \\
             \hat{\bm{\beta}}^{\text{grSCAD}/\text{grMCP}}(\lambda,\gamma)&=\underset{\beta}{\operatorname{argmin}}\{L(\bm{\beta})+\lambda \sum_{k=1}^KP_{\lambda,\gamma}^{\text{SCAD}/\text{MCP}}(||\bm{\beta}_k||_2)\} \\
            \hat{\bm{\beta}}^{\text{SGL}}(\lambda)&=\underset{\beta}{\operatorname{argmin}}\{L(\bm{\beta})+\lambda_1\sum_{k=1}^K||\bm{\beta}_k||_2+\lambda_2||\bm{\beta}||_1\}
             \end{align*}

             \subsection{Variable Clustering with PCAMIX}
             The goal of variable clustering is to group a set of variables into some homogeneous groups and is therefore variable clustering can seek for a meaningful group structures. \citet{chavent2011clustofvar} proposed a hierarchical variable clustering algorithm that can perform cluster analysis among the variables based on a PCAMIX (\citet{kiers1991simple}) method. The algorithm can be operated on variable clustering with no restriction on quantitative or qualitative variables, and is a flexible version of Principle Component Analysis (PCA). 
             
             \subsubsection{PCAMIX Calculation Algorithm}\label{sec:pcamix}
             Consider a set of $k_1$ quantitative variables and a set of $k_2$ qualitative variables, denoted as $\bm{A}$ and $\bm{B}$, and $n$ is the number of observations in $\bm{A}$ and $\bm{B}$, such that, $\bm{A}_{n\times k_1}=\{a_1,a_2,a_3,...,a_{k_1}\}$ and $\bm{B}_{n\times k_2}=\{b_1,b_2,b_3,...,b_{k_2}\}$. The detail information of variable clustering is followed on next sections. Define $\bm{D}$ is the indicator matrix of $\bm{B}$, $\Sigma$ is the diagonal matrix of each categories frequency in $\bm{B}$, and $\bm{\Phi}=\bm{I}-\frac{\bm{1}^T\bm{1}}{n}$ is the centering operator, where $\bm{I}$ is a identity matrix and $\bm{1}$ is the vector with unit $1$ entries. Thus, the algorithm of PCAMIX is defined as
             \begin{algorithm} 
            	\caption{PCAMIX} 
            	\begin{algorithmic}[1]
                	\State \textbf{input}: $\bm{A}$, $\bm{B}$, $n$ 
                	\State $\bm{A}\gets \bm{A-1}\bm{1}^T\bm{A}n^{-1}$ 
                	\Comment{Standardize $\bm{A}$}
                	\State $\bm{B}\gets \bm{\Phi}\bm{D}\bm{\Sigma}^{-\frac{1}{2}}$
                	\Comment{Standardize $\bm{B}$}
                	\State $\bm{W_k}\gets \frac{1}{\sqrt{n}}\bm{A}| \bm{B}$
                	\Comment{CONCAT $\bm{A}$ and $\bm{B}$}
                	\State Solve $\bm{W_k}=\bm{U}\bm{\Lambda}\bm{V}^T$
                	\Comment{SVD on $\bm{W_k}$}
                	\State \textbf{return}: $\bm{U},\bm{\Lambda}, \bm{V}$\Comment{A matrix of PC scores}
            	\end{algorithmic} 
            \end{algorithm}
            
            \subsubsection{Homogeneity}
            Define a partition set $K_M$ is a partition of M clusters from $k_1+k_2$ variables, such that, $K_M=\{C_1,C_2,C_3,...,C_M\}$, where $C_m$ is the $m$th cluster in $K_M$. Thus, the homogeneity within cluster $C_m$ is defined as
            \begin{align}\label{eq:2.11}
            H(C_m)=\lambda^{m}_1
            \end{align}
            where $\lambda^{m}_1$ is the first eigenvalues in $\bm{\Lambda}$ of PCAMIX (Section \ref{sec:pcamix}) applied to cluster $C_m$. The homogeneity H of partition $K_M$ is defined as is defined as sum of all the homogeneities of its clusters, such that
            \begin{align}\label{eq:2.12}
            H(K_M)=\sum_{m=1}^M H(C_m)=\sum_{m=1}^M \lambda_1^m
            \end{align}
            The homogeneity $H(C_m)$ has maximum value when all the variables within $C_m$ bring the same information.
            
            \subsubsection{Hierarchical Variable Clustering Algorithm}
            Suppose we have a dataset $D_{n\times k}$, where n is the number of observations, k is the number of variables, and $A$, $B$ are the two clusters of $D_{n\times k}$. The dissimilarity D of clusters $A$ and $B$ is defined as 
            \begin{align}\label{eq:2.13}
            D(A,B)=H(A)+H(B)-H(A\cup B)
            =\lambda_1^{A}+\lambda_1^{B}-\lambda_1^{A\cup B}
            \end{align}
            Hence, the algorithm of hierarchical variable clustering $\textbf{clust}$ can be defined as
            \begin{algorithm}
            \caption{Hierarchical Variable Clustering}
                \begin{algorithmic}[1]
                \State \textbf{input}: $D_{n\times k}$
                \State  $C_{m}\gets \{\}$, $Q\gets \langle \rangle$
                \Comment{$Q$, $c_m$: set operator}
                \While{$k>0$}\Comment{At least one cluster}
                    \State $C\gets \{\comb{k}{2}\}$ \Comment{Cases of $k$ choose 2}
                    \ForEach{$(c_1,c_2)\subseteq C$}
                        \State $d\gets D(c_1,c_2)$  \Comment{Dissimilarity of $c_1$ and $c_2$}
                        \State $Q\langle d\rangle$ $\gets (c_1,c_2)$ \Comment{Insert $d$, $(c_1,c_2)$ into $Q$}
                    \EndFor
                    \State $(c_a,c_b)\gets Q\langle \min(d)\rangle$ \Comment{Take minimum value}
                    \State $C_m\gets C_m\cup (c_a,c_b)$ \Comment{Insert $(c_a,c_b)$} into $C_m$
                    \State $C[(c_a,c_b)]\gets C[c_a\cup c_b]$ \Comment{Replace $\{c_a,c_b\}$}
                    \State $k\gets k-1$ \Comment{Now we have k-1 variables}
                \EndWhile
                \State \textbf{return}: $C_m$ \Comment{The partition of a cluster}
            \end{algorithmic}
            \end{algorithm}
            
            \subsubsection{Rand Index (RI) and Adjusted Rand Index (ARI)}
            A traditional criterion for measuring the accuracy of clustering is the Rand Index (RI), which was proposed by \citet{rand1971objective}. Consider a set of n elements $W=\{w_1,w_2,...,w_n\}$, define $X=\{x_1,x_2,...,x_S\}$ represents the true information of categories and $Y=\{y_1,y_2,...,y_{J}\}$ represents the clusters for $X$, 
            \begin{align*}
                \cup_{s=1}^{S}x_s=\cup_{j=1}^{J}y_j=W\\
                x_s\cap x_{s'}=y_j\cap y_{j'}=\varnothing \nonumber
            \end{align*}
            where $1\le s\ne s'\le S$ and $1\le j\ne j'\le J$. Suppose the True Positive (TP) is the pairs of element in $X$ and also in $Y$, and the True Negative (TN) is the pairs of element not in $X$ and also not in $Y$. The Rand Index (RI) is defined as
            \begin{align}\label{eq:2.14}
                \text{RI} = \frac{\text{TP+TN}}{\comb{n}{2}} 
            \end{align}
            where $\comb{n}{2}$ is the total combination of pairs elements, and RI $\in [0,1]$, the larger size of RI, the better performance of clustering. 
            
            Even though Rand Index (RI) can be measured the outcomes of clustering, for two stochastic partitions, the expected value of them must be greater than or equal to zero, which is not a constant value. \citet{hubert1985comparing} proposed the Adjusted Rand Index (ARI) based on  generalized hypergeometric distribution, and it is defined as 
            \begin{align}\label{eq:2.15}
                \text{ARI} = \frac{\text{RI}-\text{E(RI)}}{\max(\text{RI})-\text{E(RI)}}
            \end{align}
             Suppose $n_{ij}$ in $W$ is the number of elements that are also in $X$ and $Y$. Thus, the expression of TP+TN can be simplified as $\sum_{i,j}\comb{n_{ij}}{2}$ and $E(\sum_{i,j}\comb{n_{ij}}{2})=\frac{\sum_i\comb{n_{i.}}{2}\sum_j\comb{n_{.j}}{2}}{\comb{n}{2}}$. After some algebra works, \citet{hubert1985comparing} simplified ARI as
            \begin{align*}
                \text{ARI} = \frac{\sum_{i,j}\comb{n_{ij}}{2}-[\sum_i\comb{n_{i.}}{2}\sum_j\comb{n_{.j}}{2}]/\comb{n}{2}}{[\sum_i\comb{n_{i.}}{2}+\sum_j\comb{n_{.j}}{2}]/2-[\sum_i\comb{n_{i.}}{2}\sum_j\comb{n_{.j}}{2}]/\comb{n}{2}}
            \end{align*}
            The expected value of ARI is zero since the range of ARI lies on $[-1,1]$. Therefore, the values that can be taken on ARI has wider range than it on RI. Hence, ARI is preferred than RI to measure the clustering performance.
            
            \subsubsection{Stability of Variable Clustering}
            Suppose $m \in M$ is the optimal number of clusters from $M=\{2,3,...,p-1\}$ in a dataset $D_{n\times p}$. The optimal number of clusters $m$ is determined by the mean of ARI from $J$ times of the bootstrap with $n$ sample for each time. The algorithm of finding $m$ is defined as (Note: Suppose a function \textbf{cutree} has been produced to cut the dendrogram)
            \begin{algorithm}
            \caption{Stability of Clustering}
                \begin{algorithmic}[1]
                \State \textbf{input}: $D_{n\times p}$, $J$
                \State  $Q\gets[\enspace]$ \Comment{$Q$: An array for mean of ARI}
                \State $C_m\gets \textbf{clust}(D_{n\times p})$
                 \Comment{Initial hierarchy }
                \For{$i\gets 2,3,...,k-1$}
                    \State $q\gets[\enspace]$ \Comment{$q$: An array for ARI}
                    \For{$j\gets 1,2,3...,J$}
                        \State $C_{ij}\gets \textbf{cutree}(C_m,i)$ \Comment{Cut dendrogram}
                        \State $q[j]\gets$ $\textbf{ARI}(C_{ij},C_m)$ \Comment{Compute ARI}
                    \EndFor
                    \State $Q[i]\gets \text{Avg}(q)$
                    \Comment{Compute the avergae of ARI}
                \EndFor
                \State \textbf{return}: index of $\max(Q)$
                \end{algorithmic}
            \end{algorithm}
            
            Therefore, following the algorithms of the variable clustering, the steps of variable clustering can be clearly expressed as Figure 2, which displays the flow task of clustering. Each general step has been shown and discussed from Algorithm 1-3 in above sections.
            
            \tikzstyle{startstop} = [rectangle, rounded corners, minimum width=2cm, minimum height=0.4cm, text centered, draw=black, fill=red!30]
            \tikzstyle{process} = [rectangle, minimum width=2.5cm, minimum height=0.8cm, text centered, draw=black, fill=orange!30]
            \tikzstyle{arrow} = [thick,->,>=stealth]
            		\begin{figure}[H]
            		\centering
            		\caption{Flow Task of Variable Clustering}
            		\begin{tikzpicture}[node distance=1cm]
            		<TikZ code>
            		\node (start) [startstop] {Start: Data ($p$ clusters)};
            		\node (pro1) [process, below of=start] {Perform Hierarchical clustering};
            		\node (pro2) [process, below of=pro1,align=center] {For each possible cluster size,take $J$ bootstrap\\ samples of $n$ observations, get dendrograms};
            		\node (pro3) [process, below of=pro2,align=center] {Compare the partitions of these $J$ \\dendrograms with initial hierarchy via ARI };
            		\node (pro4) [process, below of=pro3] {Cut dendrogram based on stability};
            		\node (end) [startstop, below of=pro4] {End: The cluster output};
            		\draw [arrow](start) -- (pro1);
            		\draw [arrow](pro1) -- (pro2);
            		\draw [arrow](pro2) -- (pro3);
            		\draw [arrow](pro3) -- (pro4);
            		\draw [arrow](pro4) -- (end);
            		\end{tikzpicture}
            	\end{figure}
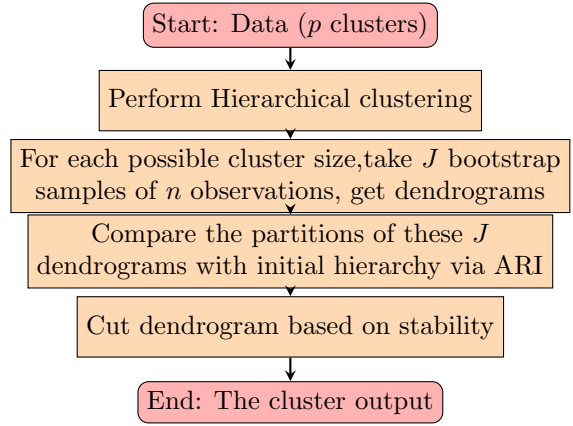
            
            \section{Two-Stage Group Variable Selection}
            We propose a two-stage approach for group variable selection in both regression and classification models. The first stage involves variable clustering via the PCAMIX hierarchical clustering method, and the second stage if the traditional group variable selection approaches. This method works well if the the number of predictors is relatively small. When the number of predictors is much larger than the number of observations, especially for ultrahigh dimensional data, the variable clustering could be computational expensive and inaccurate, especially for stability of variable clustering, which involves the bootstrap procedure. To deal with such challenge, we adjust the first stage by first using variable screening methods to keep a small subset of the predictors, then conduct the variable clustering. Specifically, we use a distance correlation based variable screening method (DC-SIS, \cite{li2012feature}) which will be introduced in more detail in Section \ref{sec:varscreen}. The following is the two-stage group variable selection algorithm. 
            \begin{algorithm}[H]
               \caption{Two-Stage Group Variable Selection}
                \begin{algorithmic}
                  \State \textbf{input:} Dataset $(X_{n\times p}, Y_{n\times 1})$
                \State \textbf{steps:}
                \State \textbf{stage 1:}
                \State \textbf{case 1:} $p\gg n$ 
                \State 1. Perform DC-SIS on dataset $(X_{n\times p}, Y_{n\times 1})$, denoted it as $X_{n\times q}$, where $q\in p$.
                \State 2. Perform the hierarchical variable clustering.
                \State 3. Choose a optimal number of clusters based on clustering stability, denoted clusters as $C_1,C_2,\cdots,C_m$.
                \State \textbf{case 2:} $p < n$ 
                \State 1. Skip step 1 in \textbf{stage 1} and repeat step 2-3.
                \State
                \State \textbf{stage 2:}
                \State 1. Run group regularization model $M$ with group structure $C_1,C_2,\cdots,C_m$.
                \State 2. $\hat{X}_{n\times l}$ is a set of selected variables, $l\in q$.
                \State \textbf{return} $\hat{X}_{n\times l}$
                \end{algorithmic}
            \end{algorithm}
            
            \subsection{Variable Screening} \label{sec:varscreen}
           In ultrahigh dimensional data, where the number of predictors are extremely large, the regularization models may not be precisely performed because of the computational expensive and stability of algorithm. Variable screening, which is also known as variable pre-selection, such as genomics data, can sufficiently reduce the variable dimension to help with the performance in terms of variable selection.
           
            \subsubsection{Sure Independence Screening (SIS)}
            Sure screening represents the variables after applying variable screening. \citet{fan2008sure} proposed a correlation learning method based on Pearson correlation, such that
            \begin{align} \label{eq:3.1}
            \rho_{x,y} = \frac{\sum_{i}^{n} (x_i-\bar{x})(y_i-\bar{y})}{\sqrt{\sum_{i}^{n} (x_i-\bar{x})}\sqrt{\sum_{i}^{n}(y_i-\bar{y})} }
            \end{align}
            named as Sure Independence Screening (SIS), to reduce the dimension of predictors space based on Pearson correlation. The general procedure of SIS is to rank the Pearson correlation between each predictors $X$ and response $Y$, then compute the minimum model size that include all active variables, denoted as $S$. The smaller value of $S$, the better performance of screening.

            \subsubsection{Distance Correlation - Sure  Independence  Screening (DC-SIS)}
            In nonlinear cases, even Pearson correlation is 0, there are no sufficient evidences to determine two random variables are independent. \citet{szekely2007measuring} introduced the distance correlation to address the deficiency of Pearson correlation. Define the distance correlation of two random vectors $u$ and $v$ denoted as $\eta_{u,v}$, such that
            \begin{align} \label{eq:3.2}
            \hat{\eta}_{u,v}=\frac{\hat{\text{dcov}}(u,v)}{\sqrt{\hat{\text{dcov}}(u,u)\hat{\text{dcov}}(v,v)}}
            \end{align}
            where $\hat{\text{dcov}}(u,v)=\hat{S_1}+\hat{S_2}-2\hat{S_3}$, and $\hat{S_1},\hat{S_2},\hat{S_3}$ are defined as
            \begin{align*}
            \hat{S_1}&=\frac{1}{n^2}\sum_{i=1}^n\sum_{j=1}^n ||u_i-u_j||_{p}||v_i-v_j||_{q}\\
            \hat{S_2}&=\frac{1}{n^2}\sum_{i=1}^n\sum_{j=1}^n ||u_i-u_j||_{p} \frac{1}{n^2}\sum_{i=1}^n\sum_{j=1}^n ||v_i-v_j||_{q}\\
            \hat{S_3}&=\frac{1}{n^2}\sum_{i=1}^n\sum_{j=1}^n\sum_{l=1}^n||u_i-u_l||_{p}||v_j-v_l||_{q}
            \end{align*}
            where $q,p$ represents the dimensions of u and v, respectively. From equation \eqref{eq:3.2}, $\hat{\text{dcov}(u,v)}=0$ if and only if two random vectors $u$ and $v$ are independent. \citet{li2012feature} proposed a distance correlation learning method that combines distance correlation and SIS, so called Distance Correlation - Sure Independence Screening (DC-SIS). The algorithm of DC-SIS are very similar to SIS, it uses distance correlation as a criterion for ranking instead of Pearson correlation. Hence, the general algorithm of variable screening is defined as 
            \begin{algorithm}
            	\caption{Variable Screening}
            	\begin{algorithmic}[1]
            		\State \textbf{input}: $X_{n\times p},Y_{n\times 1}$
            		\State $Q\gets [\enspace]$
            		\Comment{$Q$: An array}
            		\State $F\gets [f_1,f_2,f_3,...,f_p]$ \Comment{$F$: Variable name}
            		\State $d\gets k\frac{n}{log(n)}$ \Comment{$k>0$}
            		\For{$i\gets 1,2,3,...,p$} \Comment{$\bm{x}_{n\times i}\subseteq \bm{X}_{n\times p}$}
            		\State $Q[i]\gets \rho_{\bm{x}_{n\times i},\bm{Y}_{n\times 1}} \enspace \text{or} \enspace\hat{\eta}_{n\times i,\bm{Y}_{n\times 1}}$
            		\Comment{$\rho$ or $\hat{\eta}$}
            		\EndFor
            		\State $W\gets Q | F$ \Comment{CONCAT Q and F}
            		\State $W\gets$ Sort($\bm{W}$) \Comment{Sort $W$ by $Q$ with descending}
            		\State \textbf{return}: $W[:d]$ \Comment{Output top $d$th rows}
            	\end{algorithmic}
            \end{algorithm}

            \section{Simulation Studies}
            In this section, we numerically explore the expediency of two-stage variable selection method, as well as to compare the performance of different group variable selection methods. We consider five simulation settings with unequal dimensions of feature space to compare the performance of regularization models with different grouping information. In simulation studies, we generate a dataset from the linear model \eqref{eq:1.1} and logistic model  \eqref{eq:1.3} with block diagonal matrices and variables in each block are correlated. The regression noise for linear model is denoted by $\epsilon\sim N(0,\sigma^2)$, where $\sigma^2$ is chosen with signal-to-noise ratio equals 1.8 (adopted from \citet{yuan2006model}) for each regression case. We denote some notations of metric for reporting, and they are defined as 
            
            a. We consider the root mean square error (RMSE) as the prediction performance in regression problem, which is defined as 
            \begin{align}
                \text{RMSE} = \sqrt{\frac{1}{n}\sum_{i=1}^n(Y_i-\hat{Y}_i)}
            \end{align}
            where $n$ is the number of observations, $y_i$ and $\hat{y}_i$ represent the true value and predicted value of response, individually. We also consider the accuracy, sensitivity, specificity and the area under the curve (AUC) as prediction performance in classification problem.
            
            b. The another sensitivity and specificity as measure of variable selection performance are defined as 
            \begin{align}
                \text{Sensitivity}=\frac{|U\cap \hat{U}|}{|T|}\\
                \text{Specificity}=\frac{|V\cap \hat{V}|}{|T|}
            \end{align}
            where $U=\{i: \beta_i \ne 0, i=1,2,\cdots,p\}$, $\hat{U}=\{i: \hat{\beta}_i \ne 0, i=1,2,\cdots,p\}$, $V=\{i: \beta_i=0, i=1,2,\cdots,p\}$, $\hat{V}=\{i: \hat{\beta}_i=0, i=1,2,\cdots,p\}$, and $T=\{i: \beta_i, i=1,2,\cdots,p\}$.
            
            We compare each simulation with two different cases that case 1 is designed as randomly define a grouping information with equal group size while case 2 is with applying two-stage variable selection. We run $50$ bootstraps for choosing optimal ARI for stability of variable clustering and repeat each simulation 500 times through parallel computing and report the several average of prediction measures, variable selection performance and computational time from 10-fold cross validation. We use R to do the computational work for each simulation, the R packages of ``ClustOfVar'' (\citet{chavent2011clustofvar}) and ``energy" (\citet{rizzo2019package}) are applied for variable clustering and variable screening and ``grpreg" (\citet{breheny2020package}) and ``SGL" (\citet{simon2018package}) are applied for each group regularization model.
            
            \subsection{Simulation 1: Dimension $(100\times 30)$}
            We generate the ${n\times p}$ dimensional predictor $X\sim MVN(0,\Sigma_{p\times p})$ with $n=100$ and $p=30$, and $\Sigma_{p\times p}$ is a diagonal block matrix between $X_i$ and $X_j$ being $\rho^{|i-j|}$, where $\rho$ is a sequence from $0.1$ to $0.9$ by 0.05. Each block is independent with its size $S=\{3,4,4,3,4,3,4,3,2\}$. Response $Y$ is simulated from equation \eqref{eq:2.4} corresponding to the true active groups $G = \{G_1^T,G_2^T,G_4^T,G_5^T,G_6^T,G_9^T\}$, where $G_i\subseteq G$ is defined as 
            \begin{align}
                \nonumber
                G_1&= \{\beta_1,\beta_2,\beta_3\}=\{0.1,0,8\}\\ \nonumber
                G_2&= \{\beta_4,\beta_5,\beta_6,\beta_7\}=\{0.4,0.3,0.2,7\}   \\ \nonumber
                G_4&= \{\beta_{12},\beta_{13},\beta_{14}\}=\{4,5,6\}\\ \nonumber
                G_5&= \{\beta_{15},\beta_{16},\beta_{17},\beta_{18}\}=\{3,0,0.5,0\}\\ \nonumber
                G_6&= \{\beta_{19},\beta_{20},\beta_{21}\}=\{0.2,0.4,0.6\}\\ \nonumber
                G_9&= \{\beta_{29},\beta_{30}\}=\{9,10\} \nonumber
            \end{align}

         The results of simulation 1 are summarized in figure 3-5. We notice that even though model with random group structure achieves higher a sensitivity score than the model with applying two-stage variable selection, they achieve a relative low score for specificity. Hence, we observe that the group regularization model with random group structure does not correctly remove the inactive variables instead it tends to build a full model. However, the model with applying two-stage variable selection achieve a solid specificity score and an acceptable sensitivity score, where group Lasso and group MCP have highest sensitivity and specificity, respectively. Figure 2-3 are also indicated each model tends to correctly remove more inactive variables with correlation increases after applying two-stage variable selection. We also notice that the RMSE of each model decreases with correlation increases, and each model with applying two-stage variable selection achieves lower RMSE than them with random group structure. In general, the sparse group Lasso has best performance than others in terms of lowest RMSE. 
         \begin{figure}[H]
             \caption{Sensitivity of Variable Selection Performance}
                \centering
                \includegraphics[width=10cm, height=7cm]{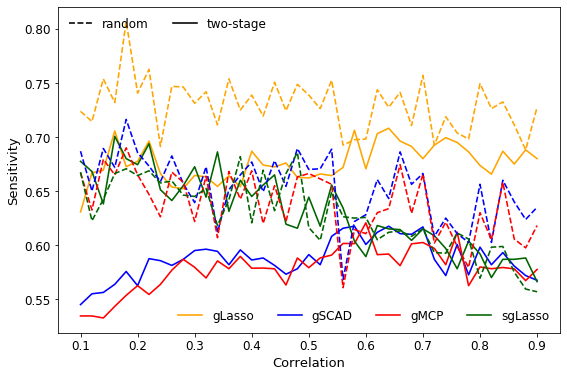}
            \end{figure}
        \begin{figure}[H]
                \centering
                \includegraphics[width=10cm, height=7cm]{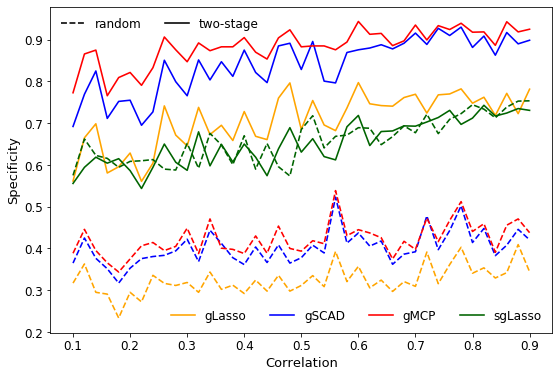}
                \caption{Specificity of Variable Selection Performance}
         \end{figure}
        \begin{figure}[H]
                \caption{RMSE with 10-fold Cross Validation }
                \centering
                \includegraphics[width=10cm, height=7cm]{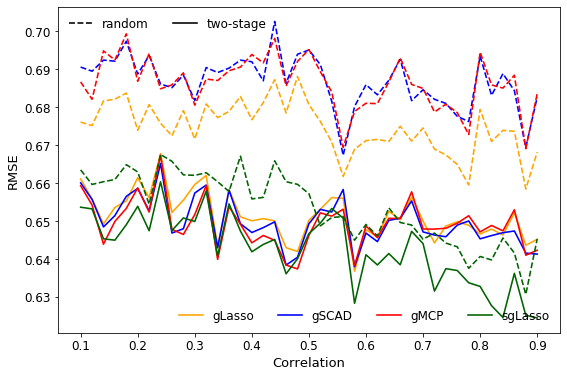}
        \end{figure}
         
        \subsection{Simulation 2: Dimension $(100\times 50)$}
        In this section, we consider to simulate a dataset involving both continuous and discrete variables. Suppose each group size $S$ is defined as $S=\{\textcolor{red}{6},\textcolor{blue}{7},\textcolor{red}{8},\textcolor{blue}{9},\textcolor{red}{10},\textcolor{blue}{10}\}$, and $\textcolor{red}{S_1}, \textcolor{blue}{S_2}
        \subseteq S$, such that \textcolor{red}{$S_1=\{6,8,10\}$}, \textcolor{blue}{$S_2=\{7,9,10\}$} and $|S_i|=\displaystyle \sum_{s\in S_i}s$. Variables  $X_{n\times |\textcolor{red}{S_1}|}$ are generated with same method as simulation 1 while variables  $X_{n\times |\textcolor{blue}{S_2}|}$ are defined as block variable $X_{i\in \textcolor{blue}{S_2}}=(U_{i\in \textcolor{blue}{S_2}}+W)/\sqrt{2}$ (refer to \citet{yuan2006model}), where $U$ and $W$ are independently generated from a standard normal distribution. We define three types of variable with repeat order of $S$, they are generated as continuous, discrete and mix (continuous and discrete) where the discrete variables are trichotomized if it is smaller than $\frac{1}{3}$ quantile, between $\frac{1}{3}$ and $\frac{2}{3}$, larger than $\frac{2}{3}$ from continuous variables and mix variables are following one by one repeat order. Response $Y$ is generated from equation \eqref{eq:2.4} with the true active groups $G=\{C_1^T,C_3^T,C_4^T,C_6^T\}$, and we define $G_i\subseteq G$ as
        \begin{align*}
            G_1&=\{\beta_1,\beta_2,\beta_3,\beta_4,\beta_5,\beta_6\}\\
                G_3&=\{\beta_{14},\beta_{16},\beta_{18},\beta_{20}\}\\
                G_4&=\{\beta_{23},\beta_{24},\beta_{25},\beta_{26},\beta_{27},\beta_{28}\}\\
                G_6&=\{\beta_{46,I(1)},\beta_{48,I(2)},\beta_{50,I(2)}\}
            \end{align*}
            where $\beta_i\sim U(-5,5)$ and $I(\cdot)$ is the indicate function represented for discrete variables. We repeat same cases as simulation 1 and report each performance measurement with typical correlations $\rho = \{0.2,0.5,0.8\}$. 
            
            The results of simulation 2 are displayed in table 1. We end with a very similar conclusion as it in simulation 1 that model with two-stage variable selection is performed better than it with random group structure because of lower RMSE. For the model with two-stage strategy, highest sensitivity and specificity score a given by the group Lasso and group MCP, respectively. The group MCP also achieves lowest RMSE comparing with others.
           
            \begin{table}[ht]
            \small
        	\centering
        	\caption{Performance Measure for Simulation 2}
    	    \begin{tabular}{ccccc}
        		\hline
        		Model & $\rho$ & RMSE & Sensitivity & Specificity  \\ \hline
        		\multicolumn{5}{l}{\textbf{case 1:} model with random equal size group} \\ \hline
        		\multirow{3}{*}{grLasso} & 0.2 & 0.87 (0.04) & 0.69 (0.08) & 0.58 (0.11)\\
        		& 0.5 & 0.87 (0.04) & 0.68 (0.10) & 0.59 (0.11) \\
        		& 0.8 & 0.87 (0.04) & 0.68 (0.10) & 0.60 (0.09)\\ \hline
        		\multirow{3}{*}{grSCAD} & 0.2 & 0.91 (0.06) & 0.65 (0.30) & 0.67 (0.15) \\
        		& 0.5 & 0.90 (0.06) & 0.66 (0.28) & 0.66 (0.15) \\
        		& 0.8 & 0.89 (0.05) & 0.74 (0.19) & 0.63 (0.10)\\ \hline
        		\multirow{3}{*}{grMCP} & 0.2 & 0.91 (0.06) & 0.64 (0.30) & 0.67 (0.16)\\
        		& 0.5 & 0.91 (0.06) & 0.65 (0.30) & 0.67 (0.15) \\
        		& 0.8 & 0.89 (0.05) & 0.73 (0.20) & 0.63 (0.11)\\ \hline
        		\multirow{3}{*}{SGL} & 0.2 & 0.83 (0.04) & 0.61 (0.10) & 0.68 (0.12) \\
        		& 0.5 & 0.81 (0.03) & 0.56 (0.11) & 0.72 (0.11) \\
        		& 0.8 & 0.81 (0.04) & 0.53 (0.10) & 0.74 (0.11) \\ \hline
        				\\
		        \hline
        		\multicolumn{5}{l}{\textbf{case 2:} model with two-stage strategy} \\ \hline
         		\multirow{3}{*}{grLasso} & 0.2 & 0.71 (0.06) & 0.71 (0.18) & 0.65 (0.18)\\
         		& 0.5 & 0.70 (0.03) & 0.71 (0.20) & 0.71 (0.18)\\
         		& 0.8 & 0.69 (0.03) & 0.80 (0.11) & 0.70 (0.16) \\ \hline
         		\multirow{3}{*}{grSCAD} & 0.2 & 0.66 (0.08) & 0.57 (0.22) & 0.79 (0.14) \\
         		& 0.5 & 0.65 (0.04) & 0.54 (0.21) & 0.83 (0.10) \\
         		& 0.8 & 0.65 (0.04) & 0.72 (0.17) & 0.78 (0.10) \\ \hline	
         		\multirow{3}{*}{grMCP} & 0.2 & 0.67 (0.08) & 0.56 (0.24) & 0.79 (0.16) \\
         		& 0.5 & 0.66 (0.04) & 0.54 (0.21) & 0.83 (0.10) \\
         		& 0.8 & 0.65 (0.04) & 0.72 (0.18) & 0.78 (0.10)\\ \hline
         		\multirow{3}{*}{SGL} & 0.2 & 0.70 (0.03) & 0.61 (0.10) & 0.68 (0.11)\\
         		& 0.5 & 0.70 (0.04) & 0.56 (0.11) & 0.72 (0.11) \\
         		& 0.8 & 0.69 (0.03) & 0.53 (0.10) & 0.74 (0.12)\\ \hline
        	\end{tabular}
    	\end{table}

    	\subsection{Simulation 3: Dimension $(100\times 150)$}
        We consider simulation 3 with same process as simulation 2 but we extend the number of group of $S$ by repeating three times and is therefore the number of dimension are higher than the number of observation, i.e, $p>n$. We set active true active groups as $G=\{G_1^T,G_3^T,G_4^T,G_6^T,G_{12}^T,G_{18}^T\}$, and $G_i\subseteq G$, which is defined as 
        \begin{align*}
            G_1&=\{\beta_1,\beta_2,\beta_3\}\\
            G_3&=\{\beta_{14},\beta_{16},\beta_{18}\}\\
            G_4&=\{\beta_{23},\beta_{24},\beta_{25},\beta_{26},\beta_{27},\beta_{28}\}\\
            G_6&=\{\beta_{46,I(1)},\beta_{48,I(2)},\beta_{50,I(2)}\}\\
            G_{12}&=\{\beta_{91},\beta_{93},\beta_{95},\beta_{100,I(1)}\}\\
            G_{18}&=\{\beta_{150,I(1)},\beta_{150,I(2)}\}
        \end{align*}
        where $\beta_i\sim U(-7,7)$ and $I(\cdot)$ represents the indicate function same as the one in simulation 2.             
        
        Table 2 displays the performance of simulation 3. We find that the models tend to blindly remove most of variables. Hence, the performance of variable selection becomes impractical in terms of selection correctness and computational time (table 5) since variable clustering is not performing well with datasets having a relatively large dimension.
        
        \begin{table}[ht]
            \small
        	\centering
        	\caption{Performance Measure for Simulation 3}
        	\begin{tabular}{ccccc}
        		\hline
        		Model & $\rho$ & RMSE & Sensitivity & Specificity  \\ \hline
        		\multicolumn{5}{l}{model with two-stage strategy} \\ \hline
        		\multirow{3}{*}{grLasso} & 0.2 & 0.98 (0.15) & 0.23 (0.12) & 0.89 (0.10) \\
        		& 0.5 & 0.95 (0.12) & 0.25 (0.12) & 0.88 (0.11)\\
        		& 0.8 & 0.94 (0.13) & 0.27 (0.11) & 0.86 (0.12)\\ \hline
        		\multirow{3}{*}{grSCAD} & 0.2 & 0.93 (0.15) & 0.11 (0.09) & 0.96 (0.03)\\
        		& 0.5 & 0.99 (0.13) & 0.17 (0.09) & 0.95 (0.04) \\
        		& 0.8 & 0.95 (0.12) & 0.13 (0.13) & 0.94 (0.05) \\ \hline
        		\multirow{3}{*}{grMCP} & 0.2 & 0.94 (0.15) & 0.08 (0.08) & 0.97 (0.04) \\
        		& 0.5 & 0.89 (0.15) & 0.14 (0.10) & 0.96 (0.04) \\
        		& 0.8 & 0.88 (0.12) & 0.11 (0.12) & 0.94 (0.05) \\ \hline
        		\multirow{3}{*}{SGL} & 0.2 & 0.85 (0.07) & 0.25 (0.08) & 0.83 (0.07)\\
        		& 0.5 & 0.84 (0.08) & 0.26 (0.07) & 0.83 (0.07) \\
        		& 0.8 & 0.84 (0.08) & 0.32 (0.08) & 0.80 (0.06) \\ \hline
        	\end{tabular}
        \end{table}
        
        \subsection{Simulation 4: Dimension $(200\times 2000)$}
        In this example, we investigate a simulation on ultrahigh dimension of variable space. The variable clustering method does not perform well for large number of variables, we consider to apply the second case in two-stage variable selection. The dataset $X_{n\times p}\sim MVN(0,\Sigma_{p\times p})$, where $n=200$ and $p=2000$ are generated with same setting as it in simulation 1 but with larger group size $S$ for each group, such that $S=\{6,7,8,9,10,10\}^{40}$ (repeat 40 times)
        The response $Y$ is generated corresponding to the active true group $G=\{G_1^T,G_3^T,G_5^T,G_6^T\}$, and $G$ is defined as
        \begin{align*}
            G_1&=\{c_1\beta_1,c_2\beta_2,c_3\beta_3,c_4\beta_4,c_5\beta_5,c_6\beta_6\}\\
            G_3&=\{c_{15}\beta_{15},c_{16}\beta_{16},c_{17}\beta_{17}\}\\
            G_5&=\{c_{31}\beta_{31},c_{32}\beta_{32},c_{33}\beta_{33}\}\\
            G_6&=\{c_{46}\beta_{46},c_{47}\beta_{47},c_{48}\beta_{48}\}
        \end{align*}
        We choose $\beta_j=(-1)^W(\eta+|Z|)$, $c_j\sim U(0.5,3)$ and $c_j\beta_j\in G$ (adopted from \citet{li2012feature}), where $W\sim Bernoulli(0,4)$, $Z\sim N(0,1)$ and $\eta=4log(n)/\sqrt{n}$.
        
        Figure 6 and 7 depict the selection performance with correlation increasing in terms of sensitivity, specificity and RMSE. The performance for applying two-stage variable selection appears much more robust than the performance with random group structure. Particularly, the specificity for each model with applying two-stage is overlapped because of excellent performance on correctly removing inactive variables, while the model with random structure displays a large heteroscedasticity on its variability. The density of sensitivity and specificity in figure 5 also indicate that there is less statistical dispersion for the model with applying two-stage than the model with random group structure. Besides, the models with applying two-stage have lower RMSE than its with random group structure, and we notice that the group Lasso with applying two-stage variable selection achieves the best overall performance comparing to others in terms of these three measurements.
        \begin{figure}[ht]
            \centering
            \includegraphics[width=12cm, height=10cm]{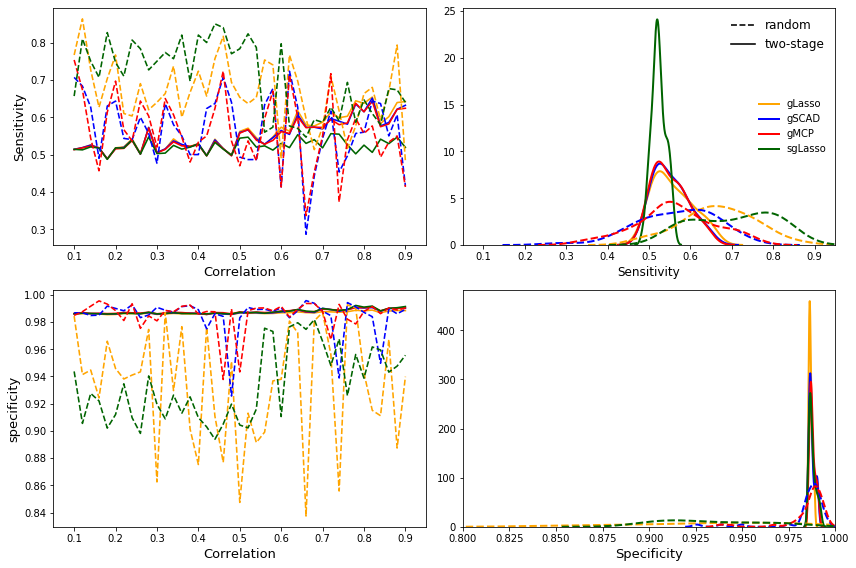}
            \caption{Performance Measurement for Simulation 4}
        \end{figure}
        \begin{figure}[H]
            \centering
            \includegraphics[width=10cm, height=6cm]{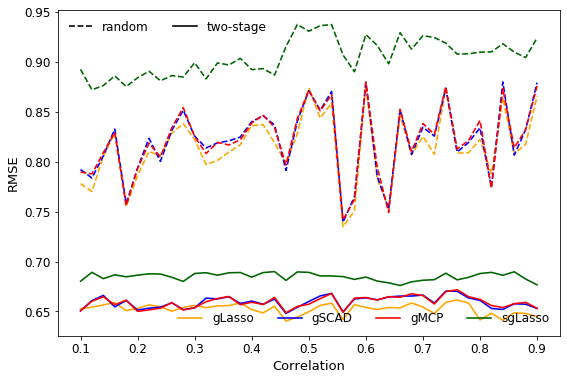}
            \caption{RMSE with 10-fold Cross Validation}
        \end{figure}
        
        \subsection{Simulation 5: Dimension $(200\times 400)$}
        In this simulation setting, we consider the performance of group regularization models in classification problem. The structure information of variable $X$ with $\rho = 0.2, 0.5$ and $0.8$, and the active groups information are determined following same setting in simulation 4 but with different group size $S$, such that $S=\{6,7,8,9,10,10\} \times 8$ (repeat 8 times). The response $Y$ (refer to \citet{sheng2013direction}) is defined as
       \begin{align*}
         P(Y=1|X)&=\frac{\exp\{g(Xc\beta)\}}{1+\exp\{g(Xc\beta)\}}\\
         g(Xc\beta)&=\frac{\exp\{5Xc\beta-2\}}{1+\exp\{5Xc\beta-3\}}-1.5
        \end{align*}
        where $c\beta$ are the active coefficients in simulation 4. 
         
        Table 3 and table 4 display the performance of simulation 5. We come up with same conclusion as it in regression type that the model with two-stage variable selection has better performance than the model with random equal size group in terms of each prediction measure in table 4 and variable selection performance in table 3. In two-stage case, group MCP achieves the best performance based on highest AUC score and excellent performance of variable selection.
        \begin{table}[H]
    	\centering
    	\small
    	\caption{Performance of Variable Selection for Simulation 5}
    	\begin{tabular}{cccc}
    		\hline
    		Model & $\rho$ & Sensitivity & Specificity   \\ \hline
    		\multicolumn{3}{l}{\textbf{case 1:} model with random equal size group} \\ \hline
    		\multirow{3}{*}{grLasso} & 0.2  & 0.48 (0.10) & 0.92 (0.06) \\
    		& 0.5 & 0.51 (0.13) & 0.89 (0.05) \\
    		& 0.8 & 0.59 (0.12) & 0.86 (0.06) \\ \hline
    		\multirow{3}{*}{grSCAD} & 0.2 & 0.46 (0.12) & 0.90 (0.07)\\
    		& 0.5 &  0.53 (0.12) & 0.90 (0.06)\\
    		& 0.8 &  0.56 (0.12) & 0.88 (0.07)\\ \hline
    		\multirow{3}{*}{grMCP} & 0.2 & 0.46 (0.12) & 0.89 (0.07)\\
    		& 0.5 & 0.53 (0.12) & 0.90 (0.06)\\
    		& 0.8 & 0.55 (0.11) & 0.89 (0.07)\\ \hline
    		\multirow{3}{*}{SGL} & 0.2 & 0.51 (0.11) & 0.85 (0.09) \\
    		& 0.5 & 0.57 (0.11) & 0.85 (0.07) \\
    		& 0.8 & 0.61 (0.11) & 0.84 (0.08) \\	\hline
    		\\
    		\hline
    		\multicolumn{3}{l}{\textbf{case 2}: model with two-stage} \\ \hline
    		\multirow{3}{*}{grLasso} & 0.2 & 0.55 (0.09) & 0.90 (0.02) \\
    		& 0.5 & 0.58 (0.09) & 0.92 (0.02) \\
    		& 0.8 & 0.65 (0.08) & 0.93 (0.02) \\ \hline
    		\multirow{3}{*}{grSCAD} & 0.2 &  0.52 (0.10) & 0.91 (0.01)\\ 
    		& 0.5 &  0.57 (0.10) & 0.92 (0.01)\\
    		& 0.8 &  0.62 (0.08) & 0.93 (0.01)\\ \hline
    		\multirow{3}{*}{grMCP} & 0.2 & 0.52 (0.08) & 0.93 (0.02)\\
    		& 0.5 &  0.58 (0.08) & 0.93 (0.02)\\
    		& 0.8 & 0.62 (0.08) & 0.94 (0.01)\\ \hline
    		\multirow{3}{*}{SGL} & 0.2 & 0.59 (0.08) & 0.87 (0.05) \\
    		& 0.5 & 0.60 (0.08) & 0.89 (0.05) \\	
    		& 0.8 & 0.65 (0.07) & 0.90 (0.05) \\ \hline
    	\end{tabular}
    \end{table}
     \begin{table}[H]
		\centering
		\small
		\caption{Prediction Performance for Simulation 5}
		\begin{threeparttable}
			\begin{tabular}{ccccc}
				\hline
				Model & Accuracy & Sensitivity & Specificity & AUC \\ \hline
				\multicolumn{4}{l}{$\rho = 0.2$} \\ \hline
				\multicolumn{5}{l}{\textbf{case 1}: model with random equal size group} \\ \hline
				grLasso & 0.56 (0.10) & 0.48 (0.11) & 0.59 (0.11) & 0.58 (0.11) \\
				grSCAD & 0.53 (0.11) & 0.50 (0.12) & 0.57 (0.11)& 0.60 (0.10) \\
				grMCP & 0.55 (0.10) & 0.49 (0.12) & 0.58 (0.12) & 0.61 (0.13)\\ 
				SGL & 0.52 (0.13) & 0.52 (0.11) & 0.53 (0.12) & 0.59 (0.11)\\
				\hline
				\multicolumn{5}{l}{\textbf{case 2}: model with two-stage} \\ \hline
				grLasso & 0.61 (0.08) & 0.56 (0.08) & 0.64 (0.08) & 0.64 (0.08) \\
				grSCAD & 0.61 (0.09) & 0.53 (0.10) & 0.68 (0.09)& 0.65 (0.08) \\
				grMCP & 0.60 (0.08) & 0.53 (0.09) & 0.68 (0.08) & 0.63 (0.08)\\ 
				SGL & 0.63 (0.07) & 0.61 (0.08) & 0.62 (0.10) & 0.62 (0.07)\\
				\hline
				\\
				\hline
				\multicolumn{5}{l}{$\rho = 0.5$} \\ \hline
				\multicolumn{5}{l}{\textbf{case 1}: model with random equal size group} \\ \hline
				grLasso & 0.55 (0.11) & 0.47 (0.12) & 0.59 (0.11) & 0.57 (0.12) \\
				grSCAD & 0.56 (0.11) & 0.49 (0.10) & 0.60 (0.12)& 0.58 (0.12) \\
				grMCP & 0.56 (0.12) & 0.49 (0.11) & 0.60 (0.11) & 0.61 (0.11)\\ 
				SGL & 0.58 (0.10) & 0.50 (0.11) & 0.61 (0.10) & 0.60 (0.12)\\
				\hline
				\multicolumn{5}{l}{\textbf{case 2}: model with two-stage} \\ \hline
				grLasso & 0.62 (0.08) & 0.57 (0.09) & 0.65 (0.08) & 0.63 (0.08) \\
				grSCAD & 0.63 (0.07) & 0.55 (0.11) & 0.68 (0.10)& 0.65 (0.07) \\
				grMCP & 0.65 (0.08) & 0.59 (0.09) & 0.70 (0.11) & 0.68 (0.09)\\ 
				SGL & 0.62 (0.07) & 0.62 (0.11) & 0.61 (0.11) & 0.66 (0.07)\\
				\hline
				\\
				\hline
				\multicolumn{5}{l}{$\rho = 0.8$} \\ \hline
				Model & Accuracy & Sensitivity & Specificity & AUC \\ \hline
				\multicolumn{5}{l}{\textbf{case 1}: model with random equal size group} \\ \hline
				grLasso & 0.56 (0.12) & 0.51 (0.13) & 0.60 (0.11) & 0.65 (0.12) \\
				grSCAD & 0.59 (0.11) & 0.48 (0.11) & 0.62 (0.12)& 0.63 (0.11) \\
				grMCP & 0.57 (0.11) & 0.47 (0.12) & 0.63 (0.12) & 0.65 (0.11)\\ 
				SGL & 0.60 (0.11) & 0.55 (0.11) & 0.61 (0.11) & 0.65 (0.10)\\
				\hline
				\multicolumn{5}{l}{\textbf{case 2}: model with two-stage} \\ \hline
				grLasso & 0.66 (0.09) & 0.61 (0.08) & 0.69 (0.08) & 0.70 (0.07) \\
				grSCAD & 0.66 (0.08) & 0.59 (0.10) & 0.71 (0.09)& 0.71 (0.09) \\
				grMCP & 0.67 (0.08) & 0.64 (0.09) & 0.72 (0.10) & 0.71 (0.09)\\ 
				SGL & 0.65 (0.07) & 0.63 (0.08) & 0.66 (0.10) & 0.69 (0.07)\\
				\hline
			\end{tabular}
			\begin{tablenotes}
				\item[1] Accuracy, Sensitivity and Specificity are measured based on 0.5 (default) decision cutoff.
			\end{tablenotes}
		\end{threeparttable}
	\end{table}
	\newpage
	
	 \subsection{Computation Time}
        In this section, we report the computing time for each simulation with with one correlation and repeating 500 times. The computation work is performed through high performance computing (HPC) in Miami Redhawk Cluster, and table 5 displays the computing time for each simulation. Miami University HPC cluster contains 26 compute nodes, each node has 24 CPU cores, and a node with 24 cores is used for the simulations. We report the user, system and elapsed time, where the user and system stands for the total time of the OS from user and elapsed stands for the real time once the running process was started. We noticed that when number of variables close to $p=150$, the computing time becomes relative long if only apply for variable clustering, instead two-stage strategy with pre-screening and variable clustering can efficiently improve the computing time for large number of variable space.
        
        \begin{table}[ht]
        	\centering
        	\small
        	\caption{Computing Time for 500 Runs (Unit: Second)}
        	\begin{threeparttable}
        	\begin{tabular}{ccccc}
        	\hline
        	Simulation & Dimension & User & System & Elapsed\\ 
        	\hline 
        	\multicolumn{5}{l}{\textbf{case 1:} model with random equal size group}\\ 
        	\hline
        	1 & $100\times 30$ & 42.8 & 0.9 & 25.6 \\
        	2 & $100\times 50$ & 102.3 & 2.3 & 116.2 \\
        	3 & $100\times 150$ & 302.2 & 2.5 & 428.1 \\
        	4 & $200\times 2000$ & 1729.3 & 2.8 & 5466.5\\
        	5 & $200\times 400$ & 709.4 & 2.6 & 2422.7 \\
        	\hline 
        	\\
		    \hline
        	\multicolumn{5}{l}{\textbf{case 2:} model with two-stage} \\
        	\hline
        	1 & $100\times 30$ & 461.3 & 1.2 &  503.4\\
        	2 & $100\times 50$ & 2022.8 & 6.4 &  2173.4\\
        	3 & $100\times 150$ & 7626.3 & 24.3 & 9415.0\\
        	4 & $200\times 2000$ & \textbf{1493.2} & \textbf{3.5} & \textbf{1637.3} \\
        	5 & $200\times 400$ & \textbf{178.0} & \textbf{1.17} & \textbf{233.8}\\
        	\hline
        	\end{tabular}
        	\begin{tablenotes}
        	\footnotesize
        	\item[1] Bold font represents the time for pre-screening + clustering.
        	\item[2] Intel Xeon Gold 6126 2.6 GHZ processors, 96 GB of memory.
        	\end{tablenotes}
        	\end{threeparttable}
        \end{table}
        
        \section{Real Data Analysis}
        \subsection{Background}
         The dataset is supported by Cincinnati Children's Hospital Medical Center and was also analyzed in \citet{paterno2010biomechanical}. The goal is to use biomechanical measures to identify will the athletes who recover the anterior cruciate ligament reconstruction (ACLR) went on suffer a second anterior cruciate Ligament (ACL) injury. The original dataset consists 118 patients and 135 variables, and we try to predict whether patient is suffered a second ACL injury, and the ratio of the response is 92 (suffered second ACL injury):26 (not suffered second ACL injury). We apply the group variable selection methods for classification.
        
        \subsection{Feature Extraction and Engineering}
        We removed some meaningless variables that are sharing the same information for the response and some variables with large missing values for the first data preprocessing step. Next, we omitted all remaining observations with missing values and encoding a few variables with discrete structure. Finally, we over-sampled dataset by applying Synthetic Minority Over-sampling Technique (SMOTE), which was proposed by \citet{chawla2002smote}, for balancing the response value. The SMOTE Algorithm can be simply defined with four steps.
        \begin{algorithm}
                \caption{Synthetic Minority Over-sampling Technique (SMOTE)}
                \begin{algorithmic}
                \State \textbf{input:} a minority class vector
                \State \textbf{steps:}
                 \State \textbf{1:} Set a sampling ratio $N$ based on imbalance ratio.
                \State \textbf{2:} For each sample $x$ in the minority class, find its $k$ nearest neighbors using the Euclidean distance    
                \State \textbf{3:} For each minority sample point $x$, randomly select a sample point from k nearest neighbors, denoted it as $\hat{x}$
                \State \textbf{4:} For each $\hat{x}$, generate a new sample point based on $x_{\text{new}}=x-\mathbb{R}_{(0,1)}(\hat{x}-x)$
                \State \textbf{return} synthetic minority class samples
                \end{algorithmic}
            \end{algorithm}
        There are 100 patients with 129 variables with the ratio of response value is 76:24 after omitting all the missing value but without using SMOTE. The number of patients are increased to 152 with 76:76 for response value if applying with SMOTE.
        
        \subsection{Modeling}
        We compared the model performance for known group structure and two-stage approach in terms of prediction and variable selection. We split data into train set (70\%) and test set (30\%), where the train set is used to find the optimal tuning parameter $\lambda$ with 10-fold cross validation, and the test set is for measuring the prediction for different models. We also compared the model performance for original dataset and the dataset with applying SMOTE on train set and report the selected variables.
        
        Table 6 displays the prediction performance in terms of accuracy, sensitivity, specificity and also the area under the curve (AUC) for based on 10-fold cross validation for each model. We notice that each prediction measurements for the dataset with SMOTE are performed better than the original dataset. Compare to the prediction measurement for case 1 and case 2, the models with two-stage approach are normally achieved higher score than the model with known group structure. In general, the group MCP has the best prediction performance based on highest score of AUC (83\%). Table 7 and table 8 show the variables that are selected by different models for the balanced response (SMOTE) and imbalanced response. We observed that there are 9 variables (with checkmark) are commonly selected by all the models with two-stage approach in table 7, while only 4 variables are commonly selected in table 8. For the model with known group structure in both table 7 and table 8, only sparse group Lasso can select partial variables from the group of biomechanical variables, since other models can only select a group of variables rather than individual variables. 
        
        \begin{table}[ht]
	    \footnotesize
	    \centering
	    \caption{Prediction Measurements for 10-fold Cross Validation}
		\begin{threeparttable}
	    \begin{tabular}{ccccc}
		\hline
		Model & Accuracy & Sensitivity & Specificity & AUC \\ \hline
		\multicolumn{4}{l}{\textbf{balanced response}: oversampling data with SMOTE} \\ \hline
		\multicolumn{4}{l}{\textbf{case 1}: known group structure } \\ \hline
		grLasso & 0.71 & 0.63 & 0.76 & 0.72 \\
		grSCAD & 0.63 & 0.64 & 0.63 & 0.70 \\
		grMCP & 0.64 & 0.61 & 0.69 & 0.69\\ 
		SGL & 0.68 & 0.64 & 0.70 & 0.71\\
		\hline
		\multicolumn{4}{l}{\textbf{case 2}: two-stage} \\ \hline
		grLasso & 0.74 & 0.69 & 0.83 & 0.79\\
		grSCAD & 0.72 & 0.67 & 0.81 & 0.78\\
		grMCP & 0.79 & 0.72 & 0.83 & 0.83\\\ 
		SGL & 0.72 & 0.71 & 0.77 & 0.76\\
		\hline
		\\
		\hline
		\multicolumn{4}{l}{\textbf{imbalanced response}: original data} \\ \hline
		\multicolumn{4}{l}{\textbf{case 1}: known group structure } \\ \hline
		grLasso & 0.75 & 0.28 & 0.91 & 0.67 \\
		grSCAD & 0.73 & 0.14 & 0.91 & 0.65\\
		grMCP & 0.73 &  0.14 & 0.91 & 0.65 \\
		SGL & 0.69  & 0.38 & 0.79 & 0.69\\ 
		\hline
		\multicolumn{4}{l}{\textbf{case 2}: two-stage} \\ \hline
		grLasso & 0.76 & 0.00 & 1.00 & 0.73\\
		grSCAD  & 0.75 & 0.01 & 0.97 & 0.71\\
		grMCP  & 0.75 & 0.01 & 0.97 & 0.71\\
		SGL  & 0.65 & 0.63 & 0.66 & 0.75 \\
		\hline
	    \end{tabular}
        \begin{tablenotes}
        \item[1] Accuracy, Sensitivity and Specificity are measured based on 0.5 (default) decision cutoff.
\end{tablenotes}
\end{threeparttable}
\end{table}
\begin{table}[H]
	\footnotesize
	\centering
	\caption{Selected Variables with Over-sampled Data}
	\begin{threeparttable}
	\begin{tabular}{ccccc}
			\hline
			\multicolumn{5}{l}{\textbf{balance response}: oversampling data} \\ \hline
			\multicolumn{5}{l}{\textbf{case 1}: known group structure } \\ \hline
			Group & grLasso & grSCAD & grMCP & SGL\\ \hline
			Injury Information (4)  & $+^*$ & & & $+_2$\\ 
			Component Score (6)  & $+^*$ &  $+^*$ &  $+^*$ & $+_3$\\
			Hop Testing (8) & & & & $+_1$\\
			Isometric Strength (7) & & & & $+_2$\\
			Isokinetic Strength (25) &  $+^*$ & & &  $+_5$\\
			Knee Laxity (7) & & & & $+_1$\\
			Postural Stability (11)  & & &  $+^*$ & $+_2$ \\
			Biomechanical Variables (61) & & & & $+_{13}$\\
			\hline
			\multicolumn{5}{l}{\textbf{case 2}: two-stage} \\ \hline
			Variable & grLasso & grSCAD & grMCP & SGL\\ \hline
			NormUnHipMomPROXIMALZIMP10 &$ \checkmark_1$ &$ \checkmark_1$ &$ \checkmark_1$&$ \checkmark_1$\\
			NormUninvHipMomentPROXIMALZIMP10&$ \checkmark_1$ &$ \checkmark_1$ &$ \checkmark_1$&$ \checkmark_1$\\\
			NormUnHipMomPROXIMALYIMP10& $ \checkmark_1$ &$ \checkmark_1$ &$ \checkmark_1$\\\
			InHipMomentPROXIMALZMINLAND&$ \checkmark_2$ &$ \checkmark_2$ &$ \checkmark_2$&$ \checkmark_2$\\
			HipUN&$ \checkmark_3$ &$ \checkmark_3$ &$ \checkmark_3$&$ \checkmark_3$\\
			DiffKneeFrontalPlaneAngYROMLAND&$ \checkmark_4$ &$ \checkmark_4$ &$ \checkmark_4$&$ \checkmark_4$\\
			DiffKneeFPAngYROMLAND&$ \checkmark_4$ &$ \checkmark_4$ &$ \checkmark_4$ &$ \checkmark_4$\\
			InKneeVelDEFAULTZIC&$ \checkmark_5$&$ \checkmark_5$ & & $ \checkmark_5$\\
			Male & $ \checkmark_6$&$ \checkmark_6$ &$ \checkmark_6$&$ \checkmark_6$\\
			InHipMomentPROXIMALXIC & $ \checkmark_7$\\
			InHipPowerDISTALXIC & $ \checkmark_7$\\
			InHipAngleDEFAULTYMAXLAND & $ \checkmark_8$&&&$ \checkmark_8$\\
			InHipAngleDEFAULTYMINLAND   & $ \checkmark_8$&&&$ \checkmark_8$\\
			InHipAngleDEFAULTYIC & $ \checkmark_8$\\
			InAnkleAngleDEFAULTYIC & $ \checkmark_9$ & $ \checkmark_9$  &$ \checkmark_9$&$ \checkmark_9$\\
			InHipMomentPROXIMALZIC  \\
			InAnkleMomentPROXIMALZMAXLAND \\
			InKneeMomentPROXIMALYMAXLAND\\
			InAnkleAngleDEFAULTXMAXLAND & &\\
			TrHavgUN & $ \checkmark_{13}$\\
			SHavgUN   & $ \checkmark_{13}$ &&&$ \checkmark_{13}$ \\
			KOOSqolINV   & $ \checkmark_{14}$ & $ \checkmark_{14}$  &$ \checkmark_{14}$&$ \checkmark_{14}$\\             
			LSI180extNorm & \\        
			\hline      
	\end{tabular}
\begin{tablenotes}
			\item[1] $+^*$ and $+_{k}$ indicate selecting all variables or number of $k$ variables, respectively. $\checkmark_{k}$ indicates selecting the variable in $k_{th}$ group.
			\item[2] $(k)$ indicates the number of variables within a group.
			\item[3] Variables are ranked by distance correlation.
		\end{tablenotes}
		\end{threeparttable}
\end{table}

\begin{table}[H]
	\footnotesize
	\centering
	\caption{Selected Variables with Original Data}
	\begin{threeparttable}
		\begin{tabular}{ccccc}
			\hline
			\multicolumn{5}{l}{\textbf{imbalance response}: original data} \\ \hline
			\multicolumn{5}{l}{\textbf{case 1}: known group structure } \\ \hline
			Group & grLasso & grSCAD & grMCP & SGL\\ \hline
			Injury Information (4)  & $+^*$ & & & $+_2$\\ 
			Component Score (6)  & $+^*$ &  $+^*$ &  $+^*$ & $+_3$\\
			Hop Testing (8) & & & & \\
			Isometric Strength (7) & & $+^*$ & & $+_3$\\
			Isokinetic Strength (25) &  $+^*$ & & &  $+_7$\\
			Knee Laxity (7) & & & & $+_1$\\
			Postural Stability (11) & & &  $+^*$ & $+_2$ \\
			Biomechanical Variables (61) & & & & $+_{17}$\\
			\hline
			\multicolumn{5}{l}{\textbf{case 2}: two-stage} \\ \hline
			Variable & grLasso & grSCAD & grMCP & SGL\\ \hline
			DiffKneeFrontalPlaneAngYROMLAND & $ \checkmark_1$ & $ \checkmark_1$ & $ \checkmark_1$&$ \checkmark_1$\\
			DiffKneeFPAngYROMLAND & $ \checkmark_1$ & $ \checkmark_1$&$ \checkmark_1$&$ \checkmark_1$\\
			InHipMomentPROXIMALYMAXLAND & $ \checkmark_1$ & $ \checkmark_1$&$ \checkmark_1$\\
			InKneeMomentPROXIMALYMAXLAND & $ \checkmark_1$ & $ \checkmark_1$&$ \checkmark_1$\\
			NormUninvHipMomentPROXIMALZIMP10 & $ \checkmark_2$&$ \checkmark_2$&$ \checkmark_2$&$ \checkmark_2$\\
			NormUnHipMomPROXIMALZIMP10 & $ \checkmark_3$& & &$ \checkmark_3$\\
			PSapsdIN & \\
			InHipMomentPROXIMALZMINLAND &\\
			HipUN & & & & $ \checkmark_5$\\
			PeakTorqueUN & & & & $ \checkmark_5$\\
			TrHavgUN &\\
			InKneeVelDEFAULTZIC &&&&$ \checkmark_6$\\
			InAnkleVelDEFAULTZIC &\\
			LSIPTNorm & $ \checkmark_7$&$\checkmark_7$&&$ \checkmark_7$\\
			ISO300extINVNorm &&&&$ \checkmark_8$\\
			KOOSqolINV  & $ \checkmark_9$&$ \checkmark_9$&$ \checkmark_9$&$ \checkmark_9$\\
			\hline
		\end{tabular}
		\begin{tablenotes}
			\item[1] $+^*$ and $+_{k}$ indicate selecting all variables or number of $k$ variables, respectively. $\checkmark_{k}$ indicates selecting the variable in $k_{th}$ group.
			\item[2] $(k)$ indicates the number of variables within a group.
			\item[3] Variables are ranked by distance correlation.
		\end{tablenotes}
	\end{threeparttable}
	\end{table}
	
	\section{Discussion and Summary}
    In this paper we compared the model performance for group Lasso, group SCAD, group MCP and sparse group Lasso. The first three group variable selection models, can only select a group of variables rather than each individual variables, while sparse group Lasso can select individual variables and group of variables, simultaneously. Hence sparse group Lasso will be more suitable for variable sparsity. Comparing to group Lasso, group SCAD and group MCP use the nonconvex penalty and achieves the Oracle properties in terms of unbiasedness and continuity. The regular group variable selection methods have the disadvantage of requiring the prior knowledge of group structure information, which is a challenging work for a raw data or high dimensional data. To deal with this challenge we introduced a two-stage variable selection. The cluster group variable selection theoretically solves the problem by seeking for a best group structure before fitting the model, but it is also impractical for large dimension of variable space because of high computational expensive and low accuracy. Hence, variable screening is necessary through pre-selecting the important variables from large variable space before variable clustering, since it improves the model performance in terms of prediction, variable sparsity and speed of computing. Our future work will be focus on revising the algorithm of variable clustering, since the original algorithm belongs to unsupervised learning, we will involve a response variable to turn the algorithm into supervised learning and compare the performance for them. We will also improve the computing speed of variable clustering through applying parallel computing.
    
    In this paper, comparing to traditional approach, the proposed two-stage variable selection method has excellent performance on variable selection in high dimensional variable space, especially for strong correlated variable structure. We discussed each model schemes and algorithm of variable clustering, and variable pre-screening from the aspect of both theory, computational studies and real data example. We displayed that two-stage approach reduce the dimension of variable space, save the computing time in variable clustering and increasing prediction accuracy and stability of variable selection. Two-stage approach complements the shortcoming of traditional group variable selection methods and researchers can apply this method and select the different penalty items by adjusting the parameters based on different research purpose.
    
    \section{Acknowledgement}
	We would like to thank for Cincinnati Children's Hospital Medical Center for providing the real dataset for analysis.
	
    \bibliographystyle{plainnat}
    \bibliography{mybib}

\end{document}